\documentclass[preprint,showpacs,preprintnumbers,amsmath,amssymb]{revtex4}
\usepackage[latin1]{inputenc}
\usepackage{graphicx}
\usepackage{dcolumn}
\usepackage{bm}
\begin{document}
\title{ First exit times and residence times for discrete random walks on finite 
lattices}
\author{S. Condamin, O. B{\' e}nichou and M. Moreau}
\affiliation{
Laboratoire de Physique Théorique de la Matière Condensée (UMR 7600), case courier 121,  Université Paris 6, 4, Place Jussieu, 75255 Paris Cedex
}%
\date{\today}

\begin{abstract}
In this paper, we derive explicit formulas for the  surface averaged first exit time  
of a discrete random walk on a finite lattice. We consider  a wide class of random walks and 
lattices, including random walks in a non-trivial potential landscape. We also
compute quantities of interest for modelling surface reactions and other dynamic processes, such as the residence time in a subvolume, the joint residence time of several particles and the number of hits on a reflecting surface.  
\end{abstract}

\pacs{05.40.Fb, 02.50.Ey}
\maketitle          

\section{Introduction}

The theory of random walks on lattices is not only a beautiful mathematical 
object, it is also useful in numerous domains of physics\cite{weiss}, including 
potential theory \cite{spitzer}, statistical field theory \cite{drouffe}, or 
biophysics \cite{goel}. Another natural application is the
diffusion of adatoms and vacancies on a crystal surface \cite{woodruff,vangastel,benichou}. 

Among the numerous issues involved in the study of such lattice random walks, one important area 
is concerned with random walks on finite lattices. There are two important reasons for that special interest. First, 
true physical systems are not infinite,  so that explicit boundary conditions have often to be taken into account in order to properly
describe situations in which confinement can be relevant. Second, exact solvable random walk  problems in bounded domains are very rare, making this   theoretical field  an important  problem in its own right\cite{montroll,mccrea,henry1,henry2,henry3,ferraro1,ferraro2,new}. 

Recently, Blanco and Fournier \cite{blanco} reported an important general 
result concerning the mean first exit time of Pearson random walks \cite{hughes} - that is  continuous space and time  
random walks,  with  a given frequency of reorientation  $\lambda$ of the direction of the constant velocity $v$ - 
in a bounded domain. They showed that the mean first exit time of a random walk
starting from the boundary of a finite domain is independant of the frequency
$\lambda$ of redirection, and is simply related to the ratio of the domain's 
volume $V$ over the surface $S$ of the domain's boundary. The corresponding 
equation is (in three dimensions): 
\begin{equation}
\left<t \right> = 4 \frac{V}{vS}
\end{equation}
where  $v$ is the speed of the walker.  
This result was extended by Mazzolo \cite{mazzolo} to the higher-order moments
of the first exit time, and by B{\' e}nichou et al. \cite{benichou05} to general 
diffusion processes in a non-uniform energy landscape. 

In this paper, we show how   these results can be extended to the important case of discrete space and time  random walks on a finite lattice. In particular,  we obtain very simple  explicit expressions for  mean exit times and  mean residence times  averaged over the surface of the considered domain, for rather general random walks. More precisely, the article is organized as follows. In section \ref{model}, we define  the model under study and the basic averages involved in the sequel. The section \ref{firstexit} is devoted to the study of first exit time moments. 
In section \ref{residence}, we generalize this approach and explicitly calculate the  mean  residence time in a subvolume of the domain. The section \ref{hits} presents an analysis of the mean number of hits of a reflecting surface. In section \ref{energy}, we generalize  the previous results to the very important case of a random walk in presence of a potential. Finally, in section \ref{jointsec},  we consider the joint residence time of several particles in 
subdomains of the lattice and derive results that can be applied, for instance
to the theory of heterogenous catalysis.    
 

\section{The model}\label{model}

\begin{figure}[h!]%
\centering\includegraphics[width = .7\linewidth]{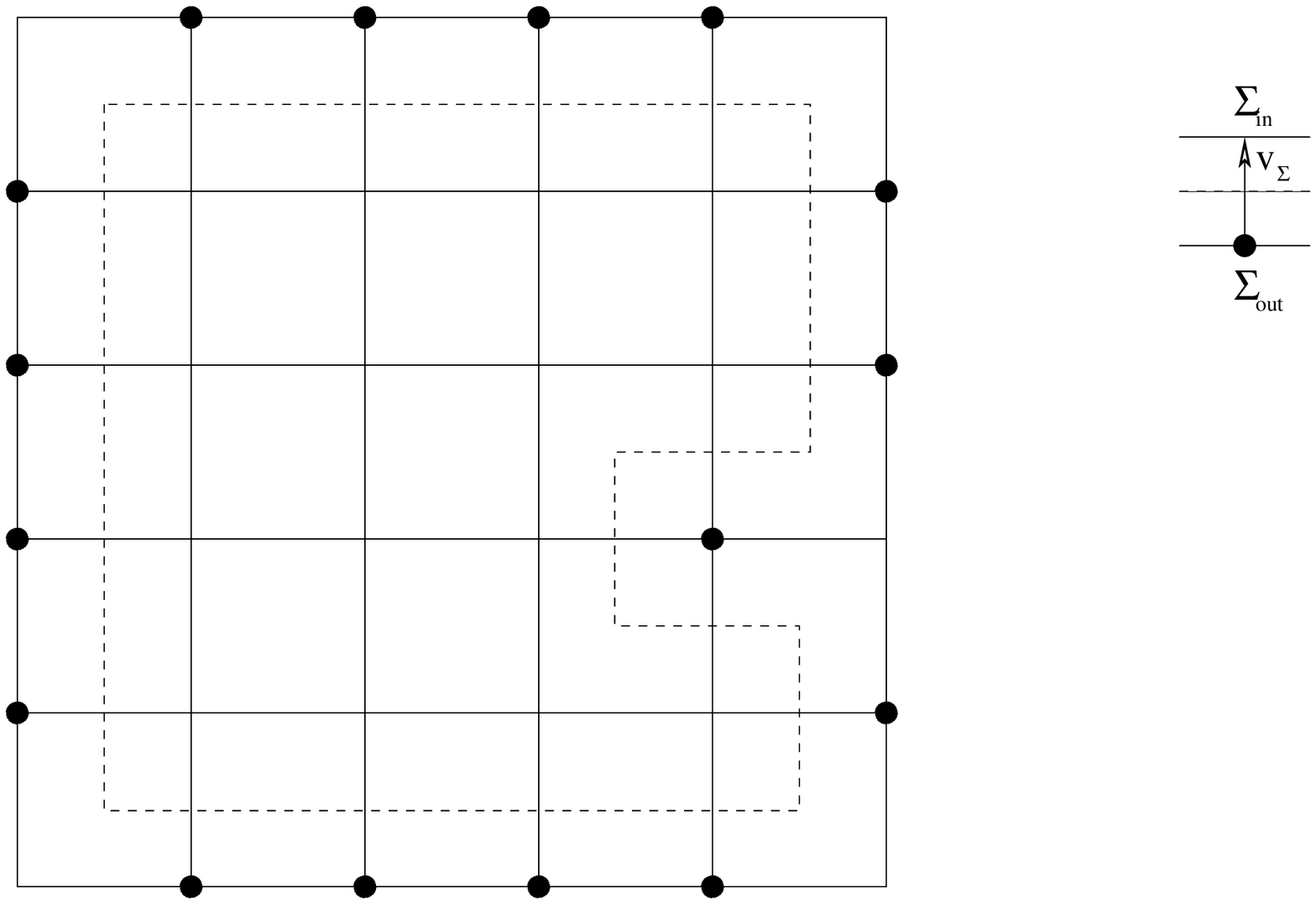}
\caption{A typical volume: The surface is the dashed line; to the right, the 
conventions for the surface}
\label{domain}
\end{figure}

Let us start with the definition of  our model. 
First, we have a lattice, which may be of any dimension or 
connectivity: for instance, we can as well apply our results to the cubic 
$3-D$ lattice than to the triangular $2-D$ lattice. 
We study the motion of a random walker (with memory: the random 
walker has a probability of switching direction $\lambda$ each time it visits a
 site; note that we can go back to the model without memory by taking 
$\lambda =1$.) 
The random walker starts from the boundary of a domain (see fig \ref{domain})

The position of the random walker is denoted by $\vec{r}(t)$; The speed of the 
random walker is $\vec{v}(t)$. The only values it can take are the difference
between the positions of two neighbours.
The precise rules of the model thus write: 
\begin{equation}
\vec{r}(t+1) = \vec{r}(t) + \vec{v}(t) 
\end{equation}
\begin{equation}
\vec{v}(t+1) = \left\{ \begin{array}{ll}
\vec{v}(t) & \mbox{with probability} \ 1-\lambda \\
\mbox{random} & \mbox{with probability} \ \lambda \\ 
\end{array}
\right. 
\end{equation}

We will also define conditional probabilities: 
$p(\vec{r}',\vec{v}',t |\vec{r}, \vec{v}) $ is the probability for the random 
walker to start from the position $\vec{r}$, with the speed $\vec{v}$ and 
arrive after a time $t$ at the position $\vec{r}'$ with the speed $\vec{v}'$. 
We also note $p(\vec{r}, \vec{v}, t)$ the probability for reaching the 
boundary after a time $t$, when starting from the point $\vec{r}$ with the 
speed $\vec{v}$; We have 
\begin{equation}
p(\vec{r}, \vec{v}, t) = \sum_{\Sigma}
p(\Sigma_{in},-\vec{v}_\Sigma,t |\vec{r}, \vec{v}) 
\label{sommebord}
\end{equation}
The sum here means a sum over all the lines which cross the boundary, and to
which we can associate a point $\Sigma_{in}$ (inside the boundary),a point $\Sigma_{out}$
(outside the boundary), and a speed $\vec{v}_\Sigma$ which points from the outside to 
the inside. (see fig.\ref{domain}) This convention means that we consider 
that the random walker has 
reached the boundary as soon as it reaches the point inside the boundary, with 
a speed pointing out. This also means that the time needed to reach the 
boundary is equivalent to the number of (not necessarily distinct) sites visited, not including 
the site the random walker begins on. 
Thus, with these probabilities, for any quantity $\varphi(t)$ 
depending on the first exit time $t$  we can define its average on $t$: 
\begin{equation}
\overline{\varphi} (\vec{r}, \vec{v})=\sum_{t=0}^{\infty} \varphi(t) 
p(\vec{r}, \vec{v}, t)
\end{equation}
For example, we can define the average exit time with simply $\varphi = t$
\begin{equation}
\overline{t}(\vec{r}, \vec{v})=\sum_{t=0}^{\infty} t p(\vec{r}, \vec{v}, t)
\end{equation}
Another such average which will be useful is the Laplace transform:
\begin{equation}
\hat{p}(\vec{r},\vec{v},s) = \sum_{t=0}^{\infty} e^{-st} p(\vec{r},\vec{v},t) 
= \overline{e^{-st}}
\end{equation}
We will then define two spatial averages of $\overline{\varphi}$: the 
first one is the surface average:
\begin{equation}
\left< \varphi \right>_{\Sigma} = \frac1S \sum_{\Sigma} \overline{\varphi} (\Sigma_{out},\vec{v}_\Sigma)
\end{equation}

 In particular,  $<t>_\Sigma $ is the mean time needed to return to the surface, 
or, alternatively, the mean number of (not necessarily different) sites visited 
between entrance and exit. 
Note that the first site, which is out of the volume, is not counted, due
to the definition of the probabilities.
$\it{S}$ is the surface, or the number of lines crossing the boundary.

There is a second useful average which may be defined, the volume average: 
\begin{equation}
\left<\varphi \right>_V = \frac1{V\sigma_D} \sum_{\vec{r} \in V, \vec{v}}\overline{\varphi}(\vec{r},\vec{v})
\end{equation}
$V$ is the volume, i.e. the number of sites inside the boundary, and 
$\sigma_D$ is 
the coordination number of the lattice. For a hypercubic lattice in dimension 
$D$, it is simply $2D$. 

An example of such a volume average is $\left< t \right>_V$, which is the
mean number of sites (still not necessarily distinct) visited by the random walk (starting from a random
point in the volume) before exiting, not including the site it starts from.
To finish this introduction, let us note a few important relations:  
\begin{equation}
\overline{\varphi}(\Sigma_{in},-\vec{v}_\Sigma)= \varphi(\Sigma_{in},-\vec{v}_\Sigma,0)
\end{equation}
This because the random walker starting at this position automatically leaves 
the volume at time 0. 
We will often have to compute terms like $\sum_{t=0}^\infty \varphi(t) 
p(\vec{r}-\vec{v},\vec{v},t+1)$. We can notice that, if $\vec{r} \in V$, then 
$p(\vec{r}-\vec{v} , \vec{v},0) = 0$. Indeed, $p(\vec{r},\vec{v})$ is equal 
to one if $\vec{r} = \Sigma_{in}$ and $\vec{v} = - \vec{v}_\Sigma$, and zero else. And, from the 
definition of $\Sigma_{in}$ and $\vec{v}_\Sigma$, we can see that there is no $\vec{r} \in V$ 
such that $\vec{r}-\vec{v} = \Sigma_{in}$, and $\vec{v} = - \vec{v}_\Sigma$.
Thus, we have: 
\begin{equation}
\sum_{t=0}^\infty \varphi(t) p(\vec{r}-\vec{v},\vec{v},t+1) = 
\sum_{t=0}^\infty \varphi(t-1) p(\vec{r}-\vec{v},\vec{v},t)
\end{equation}
Furthermore, we notice that: 
\begin{equation}
\sum_{\vec{r} \in V, \vec{v}}\overline{\varphi} (\vec{r}-\vec{v},\vec{v}) -  
\sum_{\vec{r} \in V, \vec{v}} \overline{\varphi} (\vec{r},\vec{v}) = 
\sum_{\Sigma}\overline{\varphi}(\Sigma_{out},\vec{v}_\Sigma) - \sum_{\Sigma}\overline{\varphi}(\Sigma_{in},-\vec{v}_\Sigma)
\label{surfvol2}
\end{equation}
Indeed, the first sum over the surface includes all the terms which are
in the first sum over the volume, but not in the second, whereas it is the 
opposite for the second sum over the surface. This equation slightly 
simplifies:

\begin{equation}
\sum_{\vec{r} \in V, \vec{v}}\overline{\varphi} (\vec{r}-\vec{v},\vec{v}) -  
\sum_{\vec{r} \in V, \vec{v}} \overline{\varphi} (\vec{r},\vec{v}) = 
\sum_{\Sigma}\overline{\varphi}(\Sigma_{out},\vec{v}_\Sigma) - S\varphi(0)
\end{equation}
\begin{figure}[h!]%
\centering\includegraphics[width = .5\linewidth]{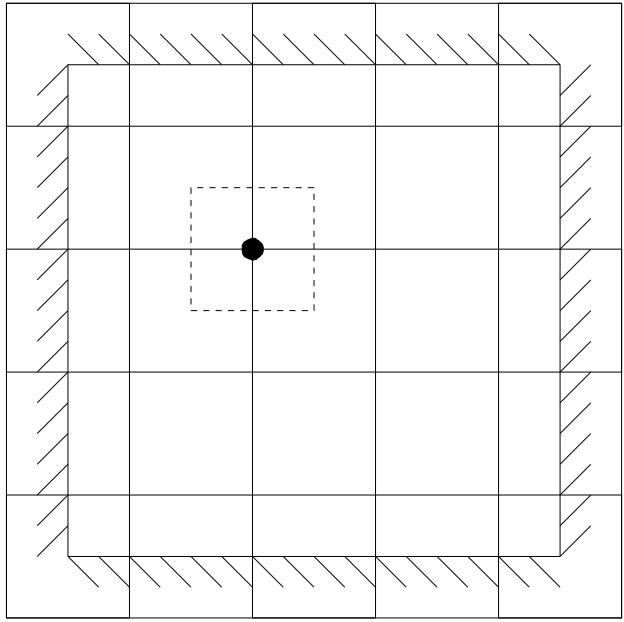}
\caption{The surface is totally reflecting; to compute the average return time
to a site, we compute the average return time to the surface surrounding 
the site, and add 1.}
\label{return}
\end{figure}%

Finally, we can easily extend the model to the case where the surface portions
may be either absorbing or reflecting. For example, it may be useful to compute
the mean return time to a site, if the boundary conditions are totally 
reflexive. (see fig \ref{return})

We denote $\Sigma$ the points of the absorbing surface and $\Sigma'$ the points of the
reflexive surface,  defined by the  reflective  boundary condition:
\begin{equation}
p(\vec{r}',\vec{v}',t|\Sigma_{in}',-\vec{v}_{\Sigma'}) = 
p(\vec{r}',\vec{v}',t|\Sigma_{out}',\vec{v}_{\Sigma'})
\end{equation}
The equation \ref{surfvol2} is thus modified the following way:
\begin{eqnarray}
\sum_{\vec{r} \in V, \vec{v}}\overline{\varphi}(\vec{r}-\vec{v},\vec{v}) -  
\sum_{\vec{r} \in V, \vec{v}}\overline{\varphi}(\vec{r},\vec{v}) & = & 
\sum_{\Sigma}\overline{\varphi}(\Sigma_{out},\vec{v}_\Sigma) - 
\sum_{\Sigma}\overline{\varphi}(\Sigma_{in},-\vec{v}_\Sigma) \\ && \nonumber
+ \sum_{\Sigma'}\overline{\varphi}(\Sigma_{out}',\vec{v}_{\Sigma'}) - 
\sum_{\Sigma'}\overline{\varphi}(\Sigma_{in}',-\vec{v}_{\Sigma'})
\end{eqnarray}
The two last terms are exactly equal, and thus we have: 
\begin{equation}
\sum_{\vec{r} \in V, \vec{v}}\overline{\varphi}(\vec{r}-\vec{v},\vec{v}) -  
\sum_{\vec{r} \in V, \vec{v}}\overline{\varphi}(\vec{r},\vec{v}) = \sum_{\Sigma}\overline{\varphi}(\Sigma_{out},\vec{v}_\Sigma) - 
S\varphi(0)
\label{surfvol}
\end{equation}
where $S$ is the surface of the absorbing portion. 

Thus, in the sequel of the article, we have to  keep in mind that all the results
 apply to a lattice where a part of the boundary is reflecting.

\section{First exit time}\label{firstexit}

 We can now proceed
to the computation of the moments of the first exit time. 
We have the following equation for the conditional probabilities: 
\begin{equation}
p(\vec{r}',\vec{v}',t+1|\vec{r}-\vec{v},\vec{v}) = 
p(\vec{r}',\vec{v}',t|\vec{r},\vec{v}) + 
\frac{\lambda}{\sigma_D} \sum_{\vec{v}''}
\left[ p(\vec{r}',\vec{v}',t|\vec{r},\vec{v''}) - 
p(\vec{r}',\vec{v}',t|\vec{r},\vec{v}) \right]
\end{equation}
This equation is simply the translation in terms of conditional probabilities 
of the rules for the behaviour of the random walker. 
We may at once sum over all the $\vec{r}'$ and $\vec{v}'$ on the boundary, as
indicated in equation (\ref{sommebord}), and we have: 
\begin{equation}
p(\vec{r}-\vec{v},\vec{v}, t+1) = 
p(\vec{r},\vec{v}, t) + \frac{\lambda}{\sigma_D} \sum_{\vec{v}''}
\left[ p(\vec{r},\vec{v''}, t) - 
p(\vec{r},\vec{v},t) \right] \label{maitre}
\end{equation}

We may now use the Laplace transforms: 

\begin{equation}
e^s \hat{p}(\vec{r}-\vec{v},\vec{v}, s) = 
\hat{p}(\vec{r},\vec{v}, s) + \frac{\lambda}{\sigma_D} \sum_{\vec{v}''}
\left[ \hat{p}(\vec{r},\vec{v''}, s) - 
\hat{p}(\vec{r},\vec{v},s)\right] 
\end{equation}

When we sum over all $\vec{r}$ and $\vec{v}$, the two last terms cancel out:  
\begin{equation}
\sum_{\vec{v}'',\vec{v}}
\overline{\varphi}(\vec{r},\vec{v''}) - 
\overline{\varphi}(\vec{r},\vec{v}) = 0
\label{annulation}
\end{equation}
and we can make use of the equation (\ref{surfvol2}) to obtain:

\begin{equation}
e^s \left[ \sum_{\Sigma} \hat{p}(\Sigma_{out},\vec{v}_\Sigma,s) - \sum_{\Sigma} \hat{p}(\Sigma_{in},-\vec{v}_\Sigma,s)\right]
= (1 - e^s) \sum_{\vec{r},\vec{v}} \hat{p}(\vec{r},\vec{v},s)
\end{equation}

Thus, we have: 

\begin{equation}
S \left( \left< e^{-st} \right>_\Sigma - 1 \right) = \left(e^{-s} - 1\right) 
\sigma_DV\left< e^{-st}\right>_V \label{laplace}
\end{equation}

If we develop each side in powers of $s$, we obtain: 
\begin{equation}
S \sum_{k=1}^\infty \frac{s^k (-1)^k}{k!} \left< t^k \right>_\Sigma = \sigma_DV 
\sum_{l=1}^\infty \frac{s^l (-1)^l}{l!} \sum_{m=0}^\infty \frac{s^m (-1)^m}{m!} \left< t^m \right>_V 
\end{equation}
If we identify, we have: 

\begin{equation}
\frac{(-1)^k}{k!} \left< t^k \right>_\Sigma = \frac{\sigma_DV}S \sum_{l=1}^k \frac1{l!(k-l)!}
\left< t^{k-l} \right>_V
\end{equation}
And finally we have the relation: 
\begin{equation}
\left< t^n \right>_\Sigma = \frac{\sigma_DV}S \sum_{m=1}^n {n \choose m} 
\left< t^{n-m} \right>_V
\label{highord}
\end{equation}

It is possible to obtain this expression directly from the evolution equation 
of $t^n$. However, the computation is slightly longer, and we will not detail 
it here. 

An important consequence of this relation is that, in the special case of the 
first moment, we have the following simple result: 

\begin{equation}
\left<t \right>_{\Sigma} = \frac{\sigma_DV}S
\label{expl1}
\end{equation}

This explicit result is quite similar to the result obtained in the 
continuous case \cite{blanco}: it has the same dependance on the surface 
and volume, but the numerical prefactor is modified. 
The simplicity of this equation makes it very easy to use. 
For instance, in the case of fig. \ref{return}, we obtain the result that the 
average return time is simply $V$. 
It depends neither on the frequency $\lambda$ nor on the shape of the volume! 

However, we must note that, for the higher-order moments, the results obtained
are different from what we have in the continuous case \cite{mazzolo}: 
in the latter case, the $n$th moment of the surfacic first exit time 
depends only on the $(n-1)$-th moment of the volumic first exit time, whereas,
on a lattice, we must take into account all the $(n-1)$ first moments of 
the volumic first exit time.

 Furthermore, we can obtain a lower bound for $\left< t \right>_V$, if we inject the value of $\left< t \right>_\Sigma$ into the equation for the second moment:

\begin{equation}
\left< t^2 \right>_\Sigma = \frac{2\sigma_DV}S \left( \left< t \right>_V + \frac12 \right)
\label{mom2}
\end{equation}
 Since we have: 
\begin{equation}
\left< t^2 \right>_\Sigma \geq \left(\left< t \right>_\Sigma \right)²
\end{equation}
we finally obtain the following bound: 
\begin{equation}
\left< t \right>_V \geq  \frac{\sigma_DV}{2S} - \frac12
\end{equation}
Note that there may be equality only in the case when every trajectory from 
surface to surface has the same length. For a square lattice, this is only 
the case for a ballistic motion (i.e. $\lambda = 0$) when the volume has a 
square shape.

\section{Residence time in a subvolume}\label{residence}

In many problems, the interesting quantity is not the time spent 
in the whole volume, but rather the time spent in a subpart of this volume.
For instance, the random walker may be a chemical reactant which may react 
exclusively on several catalytic sites. Then, the subvolume corresponds to 
the ensemble of these catalytic sites. 
Here we will only derive the first moment of the residence time in this 
subvolume $V'$, since the other moments are of limited interest, and have a 
much more complicated expression. 
What we will compute is the average residence time in $V'$, 
assuming the random walk starts and finishes at the boundary of $V$.
Thus, here, we will slightly modify the definitions of $p(\vec{r},\vec{v},t)$
and $t(\vec{r},\vec{v})$. The time will be the time spent in the 
subvolume $V'$. We thus have: 

\begin{equation}
p(\vec{r}-\vec{v},\vec{v}, t+1) = 
p(\vec{r},\vec{v}, t) + \frac{\lambda}{\sigma_D} \sum_{\vec{v}'}
\left[ p(\vec{r},\vec{v}', t) - 
p(\vec{r},\vec{v},t) \right]
\end{equation}
if $r \in V'$ .

\begin{equation}
p(\vec{r}-\vec{v},\vec{v}, t) = 
p(\vec{r},\vec{v}, t) + \frac{\lambda}{\sigma_D} \sum_{\vec{v}'}
\left[ p(\vec{r},\vec{v}', t) - 
p(\vec{r},\vec{v},t)\right]
\end{equation}
if $r \notin V'$

Thus, we have: 

\begin{equation}
\overline{t}(\vec{r}-\vec{v},\vec{v}) - 1 = 
\overline{t}(\vec{r},\vec{v}) + \frac{\lambda}{\sigma_D} \sum_{\vec{v}'}
\left[ \overline{t}(\vec{r},\vec{v}') - 
\overline{t}(\vec{r},\vec{v}) \right]
\end{equation}
if $r \in V'$
 .
\begin{equation}
\overline{t}(\vec{r}-\vec{v},\vec{v}) = 
\overline{t}(\vec{r},\vec{v}) + \frac{\lambda}{\sigma_D} \sum_{\vec{v}'}
\left[ \overline{t}(\vec{r},\vec{v}') - 
\overline{t}(\vec{r},\vec{v}) \right]
\end{equation}
if $r \notin V'$

We sum all this, which leads to: 

\begin{equation}
\sum_{\vec{v},\vec{r}} \overline{t}(\vec{r}-\vec{v},\vec{v}) = 
\sum_{\vec{v},\vec{r}} \overline{t}(\vec{r},\vec{v})
 + \frac{\lambda}{\sigma_D} \sum_{\vec{r},\vec{v},\vec{v}'} \left[
\overline{t}(\vec{r},\vec{v}') - 
\overline{t}(\vec{r},\vec{v}) \right] + \sigma_DV'
\end{equation}
The equations (\ref{surfvol}) and (\ref{annulation}) still apply here, 
and we have:
\begin{equation}
\left<t \right>_\Sigma = \frac{\sigma_DV'}{S}
\label{expl2}
\end{equation}

Thus, the residence time in a subvolume is proportionnal to its volume, and 
doesn't depend of the shape of the subvolume.
We can also say that each site is visited in average 
$\frac{\sigma_D}{S}$ times.  We check that, if the 
subvolume is in fact the whole volume, we find the same result as previously
(see equation (\ref{expl1}))

\section{Number of hits on a reflecting surface}\label{hits}

Another question which may be asked is the following: 
On average, how many times does a particle touch a portion $\Sigma''$ of the 
reflecting surface before exiting? 
If we note $t$ the number of times the particle touch the surface $\Sigma''$ 
before exiting, we have the following equations: (we note in this section 
$\Sigma'$ the rest of the reflecting surface, and $\Sigma$ is still the 
absorbing surface)
\begin{equation}
p(\vec{r}-\vec{v},\vec{v}, t) = 
p(\vec{r},\vec{v}, t) + \frac{\lambda}{\sigma_D} \sum_{\vec{v}'}
\left[ p(\vec{r},\vec{v}', t) - 
p(\vec{r},\vec{v},t) \right]
\end{equation}
and the boundary conditions: 
\begin{equation}
p(\Sigma_{in},-\vec{v}_\Sigma,t) = \delta(t,0)
\end{equation}
\begin{equation}
p(\Sigma_{in}',-\vec{v}_{\Sigma'},t) = p(\Sigma_{out}',\vec{v}_{\Sigma'},t)
\end{equation}
\begin{equation}
p(\Sigma_{in}'',-\vec{v}_{\Sigma''},t+1) = 
p(\Sigma_{out}'',\vec{v}_{\Sigma''},t)
\end{equation}

Using the Laplace transforms, we get the following equations: 

\begin{equation}
\label{base2}
\hat{p}(\vec{r}-\vec{v},\vec{v}, s) = 
\hat{p}(\vec{r},\vec{v}, s) + \frac{\lambda}{\sigma_D} \sum_{\vec{v}'}
\left[ \hat{p}(\vec{r},\vec{v}', s) - 
\hat{p}(\vec{r},\vec{v},s) \right]
\end{equation}
\begin{equation}
\hat{p}(\Sigma_{in},-\vec{v}_\Sigma,s) = 1
\end{equation}
\begin{equation}
\hat{p}(\Sigma_{in}',-\vec{v}_{\Sigma'},s) = 
\hat{p}(\Sigma_{out}',\vec{v}_{\Sigma'},s)
\end{equation}
\begin{equation}
e^s\hat{p}(\Sigma_{in}'',-\vec{v}_{\Sigma''},s) = 
\hat{p}(\Sigma_{out}'',\vec{v}_{\Sigma''},s)
\end{equation}

Here we can sum the relation (\ref{base2}) over all $\vec{r}$ and $\vec{v}$, 
and use the relations (\ref{surfvol2}) and (\ref{annulation}), which yields: 
\begin{eqnarray}
\sum_{\Sigma} \left[ \hat{p}(\Sigma_{out},\vec{v}_\Sigma) - 
\hat{p}(\Sigma_{in},-\vec{v}_\Sigma) \right]  +
\sum_{\Sigma'} \left[ \hat{p}(\Sigma_{out}',\vec{v}_{\Sigma'}) - 
\hat{p}(\Sigma_{in}',-\vec{v}_{\Sigma'}) \right] && \\ \nonumber +
\sum_{\Sigma''} \left[ \hat{p}(\Sigma_{out}'',\vec{v}_{\Sigma''}) - 
\hat{p}(\Sigma_{in}'',-\vec{v}_{\Sigma''}) \right] & = & 0
\end{eqnarray}
We define: 
\begin{equation}
\left< \varphi \right>_{\Sigma''} = \frac1{S''} \sum_{\Sigma''} 
\overline{\varphi}(\Sigma_{out}'',\vec{v}_{\Sigma''})
\end{equation}
Using the boundary conditions, we get: 
\begin{equation}
S \left( \left< e^{-st} \right>_\Sigma - 1 \right) + 
S'' \left( 1 - e^{-s} \right) \left< e^{-st} \right>_{\Sigma''} = 0
\end{equation}
If we develop this expression into powers of $s$, and identify, we finally 
get the following relation for the moments of t: 
\begin{equation}
\left< t^n \right>_\Sigma = \frac{S''}{S} \sum_{m=1}^n {n \choose m}
\left< t^{n-m} \right>_{\Sigma''}
\end{equation}
We may notice that, for the first moment, we get the simple result:
\begin{equation}
\left< t \right>_\Sigma = \frac{S''}S 
\label{expl3}
\end{equation}
This relation is again quite simple, and may easily be directly useful.

\section{Case of a non-uniform energy landscape}\label{energy}

The most simple models of random walks, which we have considered until now,
assume that all the points of the
lattice are equivalent. But, in some applications, we have to take into account
the fact that the vertices of the lattice may be at different potentials. 
For instance, in the case of a particle (vacancy, adatom) moving on a 
crystalline lattice, the presence of a inhomogeneity (typically, another
kind of atom, which would be adsorbed in the surface) may modify the effective
potential for our random walker around it. 
In the model we will introduce in this section, 
the various points of the lattice may have different energies. 
To take this particularity into account, we add to our model a reflexion 
probability. When going from the site $x$ to the site $y$, the random walker 
has a certain probability $R_{x \to y}$ to be reflected, and a probability 
$T_{x \to y}$ to be transmitted. The probabilities satisfy a detailed balance
relation, which is: 
\begin{equation}
T_{x \to y} e^{-E_x} = T_{y \to x} e^{-E_y} = A_{x,y}
\end{equation}
(We condider $kT = 1$, and scale the energies accordingly)

Thus, the law is: 

\begin{equation}
\begin{array}{l} \left\{ \begin{array}{lll} 
\vec{r}(t+1) & = & \vec{r}(t) + \vec{v}(t) \\
\vec{v}(t+1) & = & \left\{ \begin{array}{ll} 
\vec{v}(t) & \mbox{prob.} \ 1 - \lambda \\
\mbox{random} & \mbox{prob.} \ \lambda \\ 
\end{array} \right.
\\ \end{array} \right\} \mbox{prob.} \ T_{\vec{r}(t) \to \vec{r}(t) + \vec{v}(t)}
\\ \\
\left\{ \begin{array}{lll} 
\vec{r}(t+1) & = & \vec{r}(t) \\
\vec{v}(t+1) & = & \left\{ \begin{array}{ll} 
- \vec{v}(t) & \mbox{prob.} \ 1 - \lambda \\
\mbox{random} & \mbox{prob.} \ \lambda \\ 
\end{array} \right.
\\ \end{array} \right\} \mbox{prob.} \ R_{\vec{r}(t) \to \vec{r}(t) + \vec{v}(t)} \\
\end{array}
\end{equation}

As for the boundary conditions, there may be reflexions on the entrance of 
the volume. We note $A_{\Sigma_{out},\Sigma_{in}} = A_\Sigma$. 
Of course, if the particle is immediately reflected, the total time it will
have spent inside the volume will be $0$.

We also redefine the average residence time: it will be the residence time 
weighted by the Boltzmann factors of the entry sites:
\begin{equation}
\left< t \right>_\Sigma = \frac{\sum_\Sigma e^{-E_{\Sigma_{out}}}
\overline{t}(\Sigma_{out},\vec{v}_\Sigma)}{\sum_\Sigma e^{-E_{\Sigma_{out}}}}
\end{equation}

It is this average residence time we will compute here.
We have: 

\begin{eqnarray}
&& p(\vec{r}-\vec{v},\vec{v},t+1) = T_{\vec{r} -\vec{v} \to \vec{r}} \left( 
p(\vec{r},\vec{v},t) + \label{energy1}
\frac{\lambda}{\sigma_D} \sum_{\vec{v}'} \left[ p(\vec{r},\vec{v}',t) - 
p(\vec{r},\vec{v},t) \right] \right) \\
& & +  (1 - T_{\vec{r} -\vec{v} \to \vec{r}}) \left( 
p(\vec{r} - \vec{v}, -\vec{v},t) +
\frac{\lambda}{\sigma_D} \sum_{\vec{v}'} 
\left[ p(\vec{r}-\vec{v},\vec{v}',t) - p(\vec{r}-\vec{v},-\vec{v},t) \right]
\right) 
\nonumber 
\end{eqnarray}
if $\vec{r}-\vec{v} \in V$. As for the boundary conditions, we have: 

\begin{equation}
\label{energy2}
\left\{ \begin{array}{lll}
p(\Sigma_{out},\vec{v}_\Sigma,t+1) & = & T_{\Sigma_{out} \to \Sigma_{in}}
\left(  \begin{array}{l} p(\Sigma_{in},\vec{v}_\Sigma,t) \\
+ \frac{\lambda}{\sigma_D} \sum_{\vec{v}'} 
\left[p(\Sigma_{in},\vec{v}',t) - p(\Sigma_{in},\vec{v}_\Sigma,t) 
\right] \end{array} \right) \\
p(\Sigma_{out},\vec{v}_\Sigma,0) & = & R_{\Sigma_{out} \to \Sigma_{in}}
\end{array} \right.
\end{equation}

\begin{equation}
\label{energy3}
\left\{ \begin{array}{lll}
p(\Sigma_{in},-\vec{v}_\Sigma,t+1) & = & R_{\Sigma_{in} \to \Sigma_{out}}
\left( \begin{array}{l}
p(\Sigma_{in},\vec{v}_\Sigma,t) \\ + \frac{\lambda}{\sigma_D} 
\sum_{\vec{v}'} \left[ p(\Sigma_{in},\vec{v}',t) - 
p(\Sigma_{in},\vec{v}_\Sigma,t) \right] \end{array} \right) \\
p(\Sigma_{out},\vec{v}_\Sigma,0) & = & T_{\Sigma_{in} \to \Sigma_{out}}
\end{array} \right.
\end{equation}

Then, from this model, it can be shown (see Appendix \ref{calcul1}) that:
\begin{equation}
\left< t \right>_\Sigma = \sigma_D 
\frac{\sum_{\vec{r}}e^{-E_{\vec{r}}}}{\sum_{\Sigma}e^{-E_{\Sigma_{out}}}}
\label{expl4}
\end{equation}

This relation may quite easily be understood intuitively: If the energies in 
the volume are lower than the energies in the surface, the random walker will
have more difficulty exiting the volume, and, thus, will stay longer inside
the volume: the sum $\sum_{\vec{r}}e^{-E_{\vec{r}}}$ will increase. 
Inversly, if the energies in the volume are higher than the energies in the 
surface, then the particles will tend to be reflected immediately, and the 
average time spent in the volume will decrease drastically. 
Moreover, if the energy landscape is flat, the energies are identical 
everywhere in the volume and the surface, we go back to the simple result
(\ref{expl1}).

\section{Joint residence time for two particles}\label{jointsec}

We will now consider not one, but several particles diffusing independantly,
and we wish to compute joint residence times.
This is interesting in the case where we have several particles which must 
be on the same site to react\cite{benichoujoint}. For example, it may be two chemical reactants,
or a vacancy and an adatom. 
If the two particles are strongly interacting, they will react as soon as they 
are in contact. The interesting quantity here is the meeting probability $P$.
We can't compute it directly, but, given the average interaction time, we 
may have an upper bound: 
\begin{equation}
P \leq <t> 
\end{equation}
(it is interesting, since the average interaction time will genraly be much 
lower than 1.)
Otherwise, if the two particles are weakly interacting (i.e. have a small 
probability $p$ of reacting each time they meet), then the reaction 
probability will be approximatively 
\begin{equation} 
P = p <t>
\end{equation}
Thus, it is always interesting to compute the joint residence time.
So, we consider here two particles, which may either start from the bulk or 
the boundary of the volume. We define the joint residence time as the amount 
of time spent by the two particles on the same site before one of them exits.
Thus, we can define 
$q(\vec{r},\vec{v},\vec{r}',\vec{v}',t) $, the probability that the two 
particles meet $t$ times before they exit, given that the first one starts 
from the position $\vec{r}$  with the speed $\vec{v}$, and the second one 
starts from the position $\vec{r}'$ with the speed $\vec{v}'$. 
We can also define the average interaction time given the initial positions
and speeds $\overline{t}(\vec{r},\vec{v},\vec{r}',\vec{v}')$.

We have the following equation: 

\begin{eqnarray}
&& \overline{t}(\vec{r}-\vec{v},\vec{v},\vec{r}'-\vec{v}',\vec{v}') = 
(1-\lambda)^2 \overline{t}(\vec{r},\vec{v},\vec{r}',\vec{v}') + 
\frac{\lambda}{\sigma_D}(1-\lambda)  \sum_{\vec{v}''}  \nonumber
\overline{t}(\vec{r},\vec{v}'',\vec{r}',\vec{v}') \\
&& +  \frac{\lambda}{\sigma_D}(1-\lambda) \label{times}
\sum_{\vec{v}'''}  \overline{t}(\vec{r},\vec{v},\vec{r}',\vec{v}''') +  
\frac{\lambda^2}{\sigma_D^2}  \sum_{\vec{v}'',\vec{v}'''}  
\overline{t}(\vec{r},\vec{v}'',\vec{r}',\vec{v}''') + 
\delta(\vec{r},\vec{r}') 
\end{eqnarray}

(The $\delta(\vec{r},\vec{r}')$ is the Kronecker delta function, whose value 
$1$ if $\vec{r} = \vec{r}'$ and $0$ otherwise)
If we sum this over all $\vec{r},\vec{r}',\vec{v},\vec{v}'$, we notice that
the right-hand side of this equation nicely simplifies: 
\begin{equation}
\sum_{\vec{r},\vec{r}',\vec{v},\vec{v}' } 
\overline{t}(\vec{r}-\vec{v},\vec{v},\vec{r}'-\vec{v}',\vec{v}') = 
\sum_{\vec{r},\vec{r}',\vec{v},\vec{v}' } 
\overline{t}(\vec{r},\vec{v},\vec{r}',\vec{v}') + \sigma_D^2V
\label{base}
\end{equation}
The last term is simply the number of quadruplets $(\vec{r},\vec{r}',\vec{v},
\vec{v}')$ where $\vec{r} = \vec{r}'$.
We have the following relations, similar to equation (\ref{surfvol}):
\begin{eqnarray}
\sum_{\vec{r},\vec{v}} 
\overline{t}(\vec{r}-\vec{v},\vec{v},\vec{r}'-\vec{v}',\vec{v}') & = & 
\sum_{\vec{r},\vec{v}} \overline{t}(\vec{r},\vec{v},\vec{r}'-\vec{v}',\vec{v}')
+ \sum_{\Sigma} 
\overline{t}(\Sigma_{out},\vec{v}_\Sigma, \vec{r}'-\vec{v}',\vec{v}') \\ &&+ 
\sum_{\Sigma} 
\overline{t}(\Sigma_{in},-\vec{v}_\Sigma, \vec{r}'-\vec{v}',\vec{v}') \nonumber
\end{eqnarray}
The last term in this equation is zero, since if a particle is at the 
position $\Sigma_{in}$ with the speed $-\vec{v}_\Sigma$, it immediately exits.
We thus have: 
\begin{eqnarray}
\sum_{\vec{r},\vec{r}',\vec{v},\vec{v}' } 
\overline{t}(\vec{r}-\vec{v},\vec{v},\vec{r}'-\vec{v}',\vec{v}') & = &
\sum_{\vec{r},\vec{r}',\vec{v},\vec{v}' } 
\overline{t}(\vec{r},\vec{v},\vec{r}',\vec{v}') + 
\sum_{\Sigma,\vec{v}',\vec{r}'} \label{dvp}
\overline{t}(\Sigma_{out},\vec{v}_\Sigma,\vec{r}',\vec{v}') \\ && \nonumber 
+ \sum_{\vec{r},\vec{v},\Sigma'} \overline{t}(\vec{r},\vec{v},\Sigma_{out}',\vec{v}_\Sigma') + \sum_{\Sigma,\Sigma'}
\overline{t}(\Sigma_{out},\vec{v}_\Sigma,\Sigma_{out}',\vec{v}_\Sigma')
\end{eqnarray}
Finally, reporting this result in the equation (\ref{base}), we have: 

\begin{equation}
2  \sum_{\Sigma,\vec{v}',\vec{r}'} \overline{t}(\Sigma_{out},\vec{v}_\Sigma,\vec{r}',\vec{v}') + \sum_{\Sigma,\Sigma'}
\overline{t}(\Sigma_{out},\vec{v}_\Sigma,\Sigma_{out}',\vec{v}_\Sigma') =  \sigma_D^2V
\end{equation}
If we define the following average joint occupation times: 
\begin{equation}
\left< t \right>_{\Sigma V} = \frac{1}{V\sigma_DS}
\sum_{\Sigma,\vec{v}',\vec{r}'} \overline{t}(\Sigma_{out},\vec{v}_\Sigma,\vec{r}',\vec{v}')
\end{equation}
is the average joint occupation time when a particle starts from the volume 
and the other from the surface; 
\begin{equation}
\left< t \right>_{\Sigma^2} = \frac{1}{S^2}  \sum_{\Sigma,\Sigma'} \overline{t}(\Sigma_{out},\vec{v}_\Sigma,\Sigma_{out}',\vec{v}_\Sigma')
\end{equation}
is the average joint occupation time when the two particles start from the 
surface. 
Then, we have the following result: 
\begin{equation}
\sigma_D^2V = 2\sigma_DVS \left< t \right>_{\Sigma V} + S^2 
\left< t \right>_{\Sigma^2}
\label{joint}
\end{equation}

Thus, we cannot have a totally explicit result for the joint residence time; 
we just derived a relation between two different joint residence times, 
depending on where the two particles start from. However, this relation gives
immediately an useful upper bound to these joint residence times: 
\begin{equation}
\left< t \right>_{\Sigma^2} \leq \frac{\sigma_D^2V}{S^2}
\end{equation}
\begin{equation}
\left< t \right>_{\Sigma V} \leq \frac{\sigma_D}{2S}
\end{equation}
Since, in trial cases, we found out that the magnitude of the two terms in 
the relation (\ref{joint}) was similar, we have good hope that these relations
will at least give an upper bound which is of the good order of magnitude. 

It is possible to have a similar result for $n$ particles: it is a relation
between $n$ different averages (depending on the number of particles which 
start from the boundary, and the number of particles which start from the
volume.
The result (see  Appendix \ref{calcul2} for details on the computation) is the following: 
\begin{equation}
\sigma_D^nV = \sum_{m=1}^n {n \choose m} S^m(\sigma_DV)^{n-m} \left< t \right>_{\Sigma^mV^{n-m}}
\end{equation}
where $\left< t \right>_{\Sigma^mV^{n-m}}$ is the average joint residence time for 
$n$ particles, of which $m$ start from the surface and $n-m$ from the 
boundary.

\section{Conclusion}\label{conclusion}

The results obtained in this paper  significantly extend the results 
previously derived\cite{blanco,mazzolo,benichou05}. They show that the mean first exit time behaves 
differently for a discrete lattice and for a continuous media, not only
quantitatively (as for equation (\ref{expl1}), for instance), but also 
qualitatively (see the relation for higher-order moments (\ref{highord}))
It should be pointed out that we obtained  explicit exact results
 (equations (\ref{expl1}),(\ref{expl2}),(\ref{expl3}) and (\ref{expl4})), 
which is not so common  in the bounded random walks literature. Furthermore, they 
apply to all kinds of lattices and random walks: because they have   a very general
range of application, these results can be very useful when  the evolution equations cannot be solved exactly. 
In fact, it is even possible to generalize our methods  to 
irregular networks, where the connectivity can vary from one site to 
another. This may be of special interest 
for complex networks which can be found in ethology , economy, 
neural networks or social sciences. Such an extension is in progress.

\appendix

\section{Computation of the model with energies}\label{calcul1}

We can rewrite the equations (\ref{energy1}), (\ref{energy2}) and (\ref{energy3})
in terms of average time: 
\begin{eqnarray}
\label{energy4}
\overline{t}(\vec{r}-\vec{v},\vec{v}) - 1 &=& T_{\vec{r} -\vec{v} \to \vec{r}}
\left( \overline{t}(\vec{r},\vec{v}) +
\frac{\lambda}{\sigma_D} \sum_{\vec{v}'} \left[ \overline{t}(\vec{r},\vec{v}') 
- \overline{t}(\vec{r},\vec{v}) \right] \right) \\
& & +  (1 - T_{\vec{r} -\vec{v} \to \vec{r}}) \left( 
\overline{t}(\vec{r} - \vec{v}, -\vec{v}) +
\frac{\lambda}{\sigma_D} \sum_{\vec{v}'} \left[
\overline{t}(\vec{r}-\vec{v},\vec{v}') - 
\overline{t}(\vec{r}-\vec{v},-\vec{v}) \right] \right)
\nonumber \end{eqnarray}
if $\vec{r}-\vec{v} \in V$, 
If we define: 
\begin{equation}
\alpha_\Sigma = t(\Sigma_{in},\vec{v}_\Sigma) + \frac{\lambda}{\sigma_D} 
\sum_{\vec{v}'} \left[ t(\Sigma_{in},\vec{v}') - t(\Sigma_{in},\vec{v}_\Sigma) 
\right] + 1 
\end{equation}
Then we have: 
\begin{equation}
\label{Sigmaout}
t(\Sigma_{out},\vec{v}_\Sigma) = T_{\Sigma_{out} \to \Sigma_{in}} \alpha_\Sigma
\end{equation}
\begin{equation}
\label{Sigmain}
t(\Sigma_{in},-\vec{v}_\Sigma) = (1 - T_{\Sigma_{in} \to \Sigma_{out}}) \alpha_\Sigma
\end{equation}
If we sum the relation (\ref{energy4}) over all $\vec{r}$ and $\vec{v}$, 
weighted by the appropriate Boltzmann factors, we get: 
\begin{eqnarray}
& & \sum_{\vec{r},\vec{v},\vec{r}-\vec{v} \in V} e^{-E_{\vec{r}-\vec{v}}}
[ \overline{t}( \vec{r}-\vec{v}, \vec{v}) -1 ] = \nonumber \\ & &
\sum_{\vec{r},\vec{v},\vec{r}-\vec{v} \in V} A_{\vec{r},\vec{r}-\vec{v}}\left(
\overline{t}(\vec{r},\vec{v}) -\overline{t}(\vec{r} - \vec{v}, -\vec{v})
\frac{\lambda}{\sigma_D} \sum_{\vec{v}'} \left[ \begin{array}{l}
\overline{t}(\vec{r},\vec{v}') - \overline{t}(\vec{r}-\vec{v},\vec{v}') \\
+ \overline{t}(\vec{r}-\vec{v},-\vec{v}) - \overline{t}(\vec{r},\vec{v}) 
\end{array} \right] \right) \nonumber \\ & & 
+ \sum_{\vec{r},\vec{v},\vec{r}-\vec{v} \in V} e^{-E_{\vec{r}-\vec{v}}} \left( 
\overline{t}(\vec{r} - \vec{v}, -\vec{v}) + 
\frac{\lambda}{\sigma_D} \sum_{\vec{v}'} \left[
\overline{t}(\vec{r}-\vec{v},\vec{v}') - \overline{t}(\vec{r}-\vec{v},-\vec{v})
\right] \right) 
\end{eqnarray}
We will see how all these terms may transform. 
\begin{equation}
\sum_{\vec{r},\vec{v},\vec{r}-\vec{v} \in V} e^{-E_{\vec{r}-\vec{v}}}
\overline{t}( \vec{r}-\vec{v}, \vec{v}) = 
\sum_{\vec{r},\vec{v}} e^{-E_{\vec{r}}}
\overline{t}( \vec{r}, \vec{v}) - \sum_{\Sigma} e^{-E_{\Sigma_{in}}} 
\overline{t}(\Sigma_{in},-\vec{v}_\Sigma)
\end{equation}

We also have:
\begin{equation}
\sum_{\vec{r},\vec{v},\vec{r}-\vec{v} \in V} A_{\vec{r},\vec{r}-\vec{v}} \overline{t}(\vec{r},\vec{v})
= \sum_{\vec{r},\vec{v}}A_{\vec{r},\vec{r}-\vec{v}} \overline{t}(\vec{r},\vec{v}) - 
\sum_{\Sigma} \overline{t}(\Sigma_{in}, \vec{v}_\Sigma)
\end{equation}
\begin{equation}
\sum_{\vec{r},\vec{v},\vec{r}-\vec{v} \in V} A_{\vec{r},\vec{r}-\vec{v}} 
\overline{t}(\vec{r}-\vec{v},-\vec{v})
= \sum_{\vec{r}',\vec{v}'}A_{\vec{r}'-\vec{v}',\vec{r}'} \overline{t}(\vec{r}',\vec{v}') - 
\sum_{\Sigma} \overline{t}(\Sigma_{in}, \vec{v}_\Sigma)
\end{equation}
(We just take $\vec{v}'=-\vec{v}, \vec{r}'=\vec{r}-\vec{v}$). Note that these 
two expressions are identical only because of the detailed balance condition, 
which implies that the function $A$ has to be symmetrical. It is important 
since these two terms must cancel each other. 
We must also compute: 
\begin{equation}
\sum_{\vec{r},\vec{v},\vec{r}-\vec{v} \in V} A_{\vec{r},\vec{r}-\vec{v}} \overline{t}(\vec{r},\vec{v}')
= \sum_{\vec{r},\vec{v}}A_{\vec{r},\vec{r}-\vec{v}} \overline{t}(\vec{r},\vec{v}') - 
\sum_{\Sigma} \overline{t}(\Sigma_{in}, \vec{v}')
\end{equation}
\begin{equation}
\sum_{\vec{r},\vec{v},\vec{r}-\vec{v} \in V} A_{\vec{r},\vec{r}-\vec{v}}
\overline{t}(\vec{r}-\vec{v},\vec{v}')
= \sum_{\vec{r}'',\vec{v}''}A_{\vec{r}''-\vec{v}'',\vec{r}''} \overline{t}(\vec{r}'',\vec{v}') - 
\sum_{\Sigma} \overline{t}(\Sigma_{in}, \vec{v}')
\end{equation}
(We just take $\vec{v}''=-\vec{v}, \vec{r}''=\vec{r}-\vec{v}$) 
These two terms also exactly cancel each other because of the detailed balance
condition. 
We have of course similar relations with the terms in $e^{-E_{\vec{r}-\vec{v}}}$
We now use all these relations in our main equation, which yields: 
\begin{eqnarray}
&& \sum_{\vec{r},\vec{v}} e^{-E_{\vec{r}}}
\overline{t}( \vec{r}, \vec{v}) 
- \sum_{\Sigma} e^{-E_{\Sigma_{in}}} \overline{t}(\Sigma_{in},-\vec{v}_\Sigma)
 - \sum_{\vec{r},\vec{v}} e^{-E_{\vec{r}}} + \sum_\Sigma e^{-E_{\Sigma_{in}}} 
 =  \nonumber  \\ && 
\sum_{\vec{r},\vec{v}} 
e^{-E_{\vec{r}}} \overline{t}(\vec{r},\vec{v}) - 
\sum_{\Sigma}e^{-E_{\Sigma_{in}}} \overline{t}(\Sigma_{in},\vec{v}_\Sigma) 
+\frac{\lambda}{\sigma_D} \sum_{\vec{r},\vec{v},\vec{v}'} e^{-E_{\vec{r}}} 
[\overline{t}(\vec{r},\vec{v}') - \overline{t}(\vec{r},\vec{v}) ]  
\nonumber \\ && + \frac{\lambda}{\sigma_D} 
\sum_{\Sigma}\sum_{\vec{v},\vec{v}'} e^{-E_{\Sigma_{in}}} \left[
\overline{t}(\Sigma_{in},\vec{v}_\Sigma) - \overline{t}(\Sigma_{in},\vec{v}')\right] 
\end{eqnarray}
We use the relation (\ref{Sigmain}) on the left-hand side, and recognize 
$\alpha_\Sigma$ on the right-hand side, we get: 
\begin{equation}
\sum_{\Sigma} (- e^{-E_{\Sigma_{in}}} + A_\Sigma) \alpha_\Sigma - 
\sum_{\vec{r},\vec{v}}e^{-E_{\vec{r}}} + \sum_\Sigma e^{-E_{\Sigma_{in}}} 
= - \sum_\Sigma  e^{-E_{\Sigma_{in}}} (\alpha_\Sigma - 1)
\end{equation}
Since, because of the relation (\ref{Sigmaout}), we have 
\begin{equation}
\left< t \right>_\Sigma = 
\frac{\sum_\Sigma A_\Sigma \alpha_\Sigma}{\sum_\Sigma e^{-E_{\Sigma_{out}}}}
\end{equation}
We get the relation: 
\begin{equation}
\left< t \right>_\Sigma = \sigma_D 
\frac{\sum_{\vec{r}}e^{-E_{\vec{r}}}}{\sum_{\Sigma}e^{-E_{\Sigma}}}
\end{equation}

\section{Computation of the joint residence times for 
n particles}\label{calcul2}

Here we will evaluate the joint residence time for $n$ particles, i.e
the amount of time they will spend all in the same site. We will not write the
equations in the fullest detail, since it would take pages, but the calculation
is quite similar to the calculation for only two particles. 
We will define the times $\overline{t}(\vec{r}_1,\vec{v}_1,\hdots,\vec{r}_n,
\vec{v}_n)$ which are the average time the $n$ particles will spend together, 
given that the kth particle starts at the position $\vec{r}_k$, with the speed
$\vec{v}_k$. These times obey a relation quite similar to the equation 
(\ref{times}), but this relation is quite unwritable in the case of $n$ 
particles. However, we can see that the right-hand term will simplify the same
way it does for two particles, and we will get the equivalent of the relation
(\ref{base})
\begin{equation}
\sum_{\vec{r}_1,\vec{v}_1,\hdots, \vec{r}_n,\vec{v}_n } 
\overline{t}
(\vec{r}_1-\vec{v}_1,\vec{v}_1,\hdots,\vec{r}_n-\vec{v}_n,\vec{v}_n) = 
\sum_{\vec{r}_1,\vec{v}_1,\hdots,\vec{r}_n,\vec{v}_n } 
\overline{t}(\vec{r}_1,\vec{v}_1,\hdots,\vec{r}_n,\vec{v}_n) + \sigma_D^nV
\end{equation}
Again, $\sigma_D^nV$ is simply the number of 2n-uplets 
$(\vec{r}_1,\vec{v}_1,\hdots, \vec{r}_n,\vec{v}_n)$ which satisfy 
$\vec{r}_1=\hdots=\vec{r}_n$
We will also have the relation, similar to the relation (\ref{dvp}), where we 
put together the terms which are similar: 
\begin{eqnarray}
&& \sum_{\vec{r}_1,\vec{v}_1,\hdots, \vec{r}_n,\vec{v}_n } 
\overline{t}
(\vec{r}_1-\vec{v}_1,\vec{v}_1,\hdots,\vec{r}_n-\vec{v}_n,\vec{v}_n) = 
\sum_{\vec{r}_1,\vec{v}_1,\hdots,\vec{r}_n,\vec{v}_n } 
\overline{t}(\vec{r}_1,\vec{v}_1,\hdots,\vec{r}_n,\vec{v}_n) \\ &&
+ \sum_{m=1}^n {n \choose m} \sum_{\Sigma^{(1)},\hdots,\Sigma^{(m)}} 
\sum_{\vec{r}_{m+1},\vec{v}_{m+1},\hdots,\vec{r}_n,\vec{v}_n}
\overline{t}
(\Sigma_{out}^{(1)},\vec{v}_\Sigma^{(1)}, \hdots, \Sigma_{out}^{(m)},
\vec{v}_\Sigma^{(m)},\vec{r}_{m+1},
\vec{v}_{m+1},\hdots,\vec{r}_n,\vec{v}_n) \nonumber
\end{eqnarray}
The average joint residence times are defined by: 
\begin{eqnarray}
&& \left< t \right>_{\Sigma^mV^{n-m}} = \frac{1}{S^m(\sigma_DV)^{n-m}} \\
&& \sum_{\Sigma^{(1)},\hdots,\Sigma^{(m)}} 
\sum_{\vec{r}_{m+1},\vec{v}_{m+1},\hdots,\vec{r}_n,\vec{v}_n} 
\overline{t}
(\Sigma_{out}^{(1)},\vec{v}_\Sigma^{(1)}, \hdots, \Sigma_{out}^{(m)},
\vec{v}_\Sigma^{(m)},\vec{r}_{m+1}, \vec{v}_{m+1},\hdots,\vec{r}_n,\vec{v}_n)
\nonumber
\end{eqnarray}
We thus get the relation: 
\begin{equation}
\sigma_D^nV = \sum_{m=1}^n {n \choose m} S^m(\sigma_DV)^{n-m} 
\left< t \right>_{\Sigma^mV^{n-m}}
\end{equation}


\end{document}